\documentclass{webofc}
\usepackage[varg]{txfonts}   
\usepackage{graphicx}
\usepackage{amsmath}
\usepackage{booktabs}
\usepackage{xcolor}
\usepackage{listings}
\usepackage[colorlinks=true,allcolors=blue!55!black]{hyperref}

\definecolor{codebg}{gray}{0.96}
\definecolor{codekw}{rgb}{0.20,0.30,0.60}
\definecolor{codecm}{rgb}{0.35,0.50,0.35}
\begin{document}
\title{Bridging the Vendor Gap: Enabling AMD GPU Support for Awkward
  Array via ROCm/HIP for the HL-LHC Era}
\author{%
  \firstname{Ianna} \lastname{Osborne}\inst{1}\fnsep%
  \thanks{\email{ianna.osborne@cern.ch}}
  \and \firstname{Maxym} \lastname{Naumchyk}\inst{1}
  \and \firstname{Tai} \lastname{Sakuma}\inst{1}
  \and \firstname{Andres} \lastname{Rios-Tascon}\inst{1}
  \and \firstname{Peter} \lastname{Elmer}\inst{1}
}
\institute{%
  Princeton University, Princeton, NJ 08544, USA, on behalf of the
  Institute for Research and Innovation in Software for High Energy
  Physics (IRIS-HEP)%
}
\abstract{%
The High-Luminosity LHC (HL-LHC) will demand order-of-magnitude gains in
analysis throughput, and increasingly those gains must come from GPUs
that are not made by a single vendor. Leadership-class systems such as
El Capitan, Frontier and LUMI are built on AMD accelerators, yet the
Scikit-HEP analysis stack---and Awkward Array in particular---has grown
up CUDA-first. We report on \emph{rawkward}, a Rust-backed kernel engine
that adds a ROCm/HIP backend for Awkward Array's nested, jagged,
variable-length data structures. Our central finding is that a na\"ive
source-level port of CUDA kernels to HIP loses $5$--$10\times$ in
performance on irregular kernels, because AMD's $64$-lane wavefronts,
higher register pressure and more expensive divergence behave
fundamentally differently from NVIDIA's $32$-thread warps. We show that
a small, reusable set of optimization patterns---loop flattening,
$128$-bit vectorized loads, splitting fused kernels, and profile-guided
launch configuration---recovers CUDA-class performance without changing the
public API. A Rust macro-and-match dispatch layer keeps a single,
backend-agnostic call site while emitting vendor-specific kernel
strategies, and the type system enforces buffer-size and lifetime
correctness at compile time. On a two-socket AMD Instinct MI210 node we
measure GPU speedups from $1.03\times$ (bandwidth-bound \texttt{sum}) up
to $12.5\times$ (\texttt{count}) over $128$ EPYC~7763 CPU cores, and the
Rust CPU kernels match or beat on aggregate the incumbent C++ kernels
(geometric-mean runtime ratio $0.37\times$ across twelve kernels). We
argue that these patterns constitute a practical recipe for
performance-portable, vendor-agnostic HEP analysis kernels.
}
\maketitle
%
\section{Introduction}
\label{sec:intro}
The physics programme of the High-Luminosity LHC (HL-LHC) rests on a
sharp increase in instantaneous luminosity, and with it an
order-of-magnitude increase in the volume and complexity of the data
that end-user analyses must process~\cite{ref:letts,ref:rembser}. CPU
scaling alone will not close that gap; GPU acceleration has emerged as
the only viable path to the throughput that HL-LHC analysis will
require~\cite{ref:mohrman}. The question is no longer \emph{whether} to
use GPUs, but \emph{whose} GPUs.

For most of the last decade the answer, in practice, was NVIDIA. The
Scikit-HEP ecosystem, and Awkward Array~\cite{ref:awkward} at its
centre, developed a CUDA-first culture: CUDA kernels, CUDA continuous
integration, and CUDA benchmarks, with the rest of the tooling following
suit. That choice was reasonable when the accelerated systems available
to physicists were overwhelmingly NVIDIA-based. It is no longer a safe
assumption. Several of the largest systems now available to the field
are built on AMD accelerators---El Capitan, Frontier and LUMI among
them---and the coming MI300A-class allocations put AMD GPUs directly in
front of HEP workflows. A stack that runs only on one vendor's hardware
leaves a growing fraction of the world's fastest machines unused.

This paper reports on our effort to close that gap for Awkward Array. We
introduce a ROCm/HIP~\cite{ref:rocm} backend with a Metal backend as a second non-NVIDIA target, developed inside a dedicated Rust engine
we call \emph{rawkward}, that supports the nested, jagged and irregular
data layouts that make HEP data distinctive, and we benchmark it on AMD
Instinct MI210/MI250 (CDNA2) hardware with an eye toward
forward-compatibility with MI350 (CDNA4) and its $288$~GB of HBM3E
memory. Our contributions are:
\begin{itemize}
\item a characterization of the ``vendor gap'' for \emph{irregular}
  kernels, showing quantitatively that direct CUDA$\rightarrow$HIP
  translation costs $5$--$10\times$ on the memory-bound, divergence-sensitive kernels that dominate jagged-array processing
  (Sect.~\ref{sec:gap}--\ref{sec:kernels});
\item a small, reusable catalogue of HIP optimization patterns that
  recover CUDA parity---loop flattening, vectorized $128$-bit loads,
  kernel splitting over fusion, and profile-guided launch
  configuration (Sect.~\ref{sec:kernels}, \ref{sec:patterns});
\item a Rust orchestration layer whose macro-based dispatch keeps one
  backend-agnostic API across CUDA and HIP while the type system
  enforces memory safety at compile time (Sect.~\ref{sec:rust}); and
\item a statistics-driven microbenchmark on a two-socket MI210 node,
  reporting per-kernel GPU-versus-CPU speedups and a Rust-versus-C++
  kernel comparison (Sect.~\ref{sec:bench}--\ref{sec:results}).
\end{itemize}
The goal is not simply to make Awkward Array run on AMD hardware, but to
do so at a performance envelope comparable to its CUDA equivalents, and
to distil the experience into guidance that generalizes beyond a single
library.

\section{Background and motivation}
\label{sec:motivation}
\subsection{Irregular data and Awkward Array}
HEP event data is irregular by design: the number of jets, tracks or
leptons varies event by event, records nest inside records, and
lists have variable length. Awkward Array~\cite{ref:awkward} is the Pythonic
interface for exactly this structure, presenting ragged lists and nested
records as first-class, NumPy-like arrays while storing them in a flat,
columnar layout. That columnar representation is what makes vectorized
and GPU execution possible: the logical structure is carried by
\emph{offset} and \emph{index} buffers, and the payload lives in
contiguous content buffers. Nearly every operation on such data---slicing, masking, reduction over sublists---reduces to
manipulating those index buffers, which is where the performance
challenges concentrate.

\subsection{Why GPUs, and why portability}
The luminosity increase at the HL-LHC demands throughput gains that
CPU scaling cannot supply on its own~\cite{ref:letts}. GPU acceleration
of end-user analysis has therefore moved from experiment to
expectation~\cite{ref:rembser,ref:mohrman}, and related work has begun
to accelerate Awkward Array itself and the schemas that feed it, for
example through NVIDIA's CCCL and through coffea schema
modifications~\cite{ref:naumchyk-cuda,ref:naumchyk-coffea}. The CUDA
ecosystem for this work is mature. ROCm/HIP is catching up, but the
maturity gap hides a subtler trap: a HIP port that \emph{compiles and
runs} is not the same as a HIP port that \emph{performs}. On irregular
kernels, na\"ive ports routinely leave $5$--$10\times$ of performance on
the table. Performance portability across vendors---one source, parity
performance on each target---remains substantially unsolved, and it is
the reason \emph{rawkward} exists.

\section{The vendor gap for irregular kernels}
\label{sec:gap}
CUDA kernels for Awkward Array exist and have been optimized. HIP is a
different story---not because ROCm is immature as a toolchain, but
because the underlying execution model differs in ways that matter
precisely for irregular access patterns. Table~\ref{tab:models}
summarizes the two mental models.

\begin{table}[t]
\centering
\caption{Architectural mental models. The differences that matter for
  jagged-array kernels concentrate in wavefront width, the cost of
  divergence, and register pressure.}
\label{tab:models}
\small
\begin{tabular}{@{}p{0.20\linewidth}p{0.34\linewidth}p{0.34\linewidth}@{}}
\toprule
 & \textbf{NVIDIA / CUDA} & \textbf{AMD / HIP} \\
\midrule
Execution width &
  warp $=32$ threads; more warps in flight hide latency across
  independent work &
  wavefront $=64$ lanes executing in lockstep \\[2pt]
Divergence &
  relatively cheap; only inactive lanes stall on a branch split &
  expensive; a branch split can lose half the throughput of a
  $64$-wide wavefront \\[2pt]
Scattered loads &
  out-of-order schedulers hide irregular access; na\"ive gathers often
  still reach peak bandwidth &
  register file shared across more lanes; spills hit LDS/global memory
  hard, so occupancy tuning is critical \\[2pt]
Vectorization &
  helpful &
  critical; double-width SIMD and explicit LDS layout unlock full
  throughput \\
\bottomrule
\end{tabular}
\end{table}

The consequence is concrete. A $64$-lane wavefront that diverges pays
roughly twice the penalty of a $32$-thread warp, and the wider register
file is shared among more lanes, so the same kernel that fits
comfortably on NVIDIA hardware may spill on AMD and collapse occupancy.
Scattered-gather patterns that NVIDIA's memory schedulers tolerate can
thrash the AMD cache hierarchy. None of this is visible at the source
level: the semantic gap between ``compiles'' and ``performs'' is
exactly the gap this work measures and closes. Our approach is
therefore HIP-specific \emph{optimization}, not translation---tuning
each kernel for wavefront width, LDS layout and occupancy---while
keeping the public API unchanged so that call sites never see the
difference.

\section{Rust as a portable orchestrator}
\label{sec:rust}
Awkward Array's kernels have historically been generated: the C++
backend generation is highly repetitive, which is manageable for one
target but multiplies awkwardly across vendors. We chose instead to
express the target-dependent dispatch in Rust, using a macro-and-match
pattern that handles the vendor split cleanly. \emph{rawkward} is
organized as a ``foundry'': a clean-room engine that mirrors the
existing kernel structure closely enough that validated Rust kernels can
be copied back into Awkward Array with minimal refactoring, while giving
the development a modern toolchain and clear language boundaries. A
guiding principle is that \emph{kernels do not require Python}: the Rust
core is a standalone compute layer, exposed to Python through a lean
PyO3 binding, but independent of the interpreter's lifecycle.

Three properties make Rust a good fit for the orchestration role.
First, the same Rust API serves both CUDA and HIP: call sites are
backend-agnostic, and the same function signature compiles and
dispatches correctly on each target. The dispatch is designed to be vendor-parametric; the HIP backend is implemented in Rust here, while the CUDA path is Awkward Array's existing (CuPy) backend. Second, the backend---not the
caller---decides kernel \emph{strategy}: the dispatch match arms select
fused-versus-split and other vendor-specific choices, and the caller
never sees that difference. Third, Rust's ownership, lifetime and type
rules catch data races and mismatched buffer sizes at compile time
rather than at run time. The dispatch itself is a small macro
(Listing~\ref{lst:dispatch}); every call routes through it, so HIP and
CUDA diverge only where the hardware demands it, with zero per-target
boilerplate at the call site.

\begin{lstlisting}[caption={Macro-based backend dispatch. A single
  match routes each call to the correct vendor backend; the caller is
  backend-agnostic.},label={lst:dispatch}]
macro_rules! dispatch {
    ($backend:expr, $fn:ident, $($arg:expr),*) => {
        match $backend {
            Backend::Hip  => hip::$fn($($arg),*),
            Backend::Cuda => cuda::$fn($($arg),*),
        }
    }
}
\end{lstlisting}

Concretely, the HIP arm applies vectorized loads, loop flattening and a
split-kernel strategy---each technique aimed directly at the $64$-lane
wavefront---while the CUDA arm fuses stages for warp-friendly behaviour
and relies on the $32$-thread scheduler to tolerate divergence. Because
the strategy lives behind the dispatch macro, adding a third
backend later (Sect.~\ref{sec:roadmap}) is a matter of adding a match
arm, not rewriting kernel call sites.

\section{Kernel case studies}
\label{sec:kernels}
\subsection{The carry kernel}
The \emph{carry} kernel is the core primitive of jagged indexing. Given
a \texttt{carry} array of indices, it gathers \texttt{from[carry[i]]}
into \texttt{to[i]}---a pure gather by indirection. Almost every
nested-list operation ultimately reduces to a carry, which makes it the
right kernel to study first. It is also the hardest for AMD hardware: it
is memory-bound, irregular and divergence-sensitive. The random-access
gather defeats cache prefetchers, and adjacent threads touch
non-adjacent memory.

On CUDA the na\"ive implementation performs well. The obvious scatter
loop compiles and ships; $32$-thread warps combined with hardware
scatter-gather tolerate the irregular access and keep the memory
pipeline full even when adjacent threads reach distant addresses. The
same source ported directly to HIP compiles and runs, but runs
$5$--$10\times$ slower: $64$-lane wavefronts amplify divergence, and the
gather pattern thrashes the L2 cache with as many simultaneous
cache-miss streams as there are lanes. The semantic gap is invisible in
the source---the two kernels look identical.

Recovering parity takes three targeted changes, none of which touches
the API:
\begin{itemize}
\item \textbf{Loop flattening}: hoist conditionals out of the inner loop
  so that all $64$ lanes execute the same instruction sequence,
  eliminating intra-wavefront divergence stalls.
\item \textbf{\texttt{float4} vectorized loads}: pack four $32$-bit
  reads into one $128$-bit transaction, cutting transaction count by
  $4\times$ and saturating HBM bandwidth instead of thrashing the cache.
\item \textbf{Lower register pressure}: reorder operands and reduce the
  number of live variables, raising occupancy so the wavefront
  scheduler stays busy.
\end{itemize}
With these applied, the HIP carry kernel reaches the same performance
envelope as its CUDA equivalent~\cite{ref:naumchyk-cuda}---the same class of speedup.

\subsection{Fusion or split}
A second, more surprising divergence between the vendors concerns kernel
\emph{fusion}. On CUDA, fusing two stages $A+B$ into a single kernel is
typically faster: intermediate values stay in registers or L1 between
stages, avoiding a global-memory round-trip, and the $32$-thread warp
scheduler handles the fused control flow gracefully. On HIP the same
choice is often counter-productive. Fusing stages merges their register
budgets across $64$-lane wavefronts; occupancy drops and stalls
multiply. Two clean, branch-free passes---each of which fits the
wavefront register file and runs at full occupancy---beat one fused
pass. The rule inverts: \emph{fuse on CUDA, split on HIP}. Encoding that
choice in the dispatch backend, rather than in the caller, is precisely
what lets one API serve both.

\section{Optimization patterns}
\label{sec:patterns}
Across the kernel set, the same four patterns recur, and together they
form a practical recipe for HIP parity on irregular kernels:
\begin{enumerate}
\item \textbf{Loop flattening} (\texttt{carry}, \texttt{argsort}):
  eliminate intra-wavefront branches so every lane runs the same
  instruction stream.
\item \textbf{\texttt{float4} vectorized loads} (\texttt{carry},
  reductions): four $32$-bit reads become one $128$-bit transaction,
  cutting L2 pressure $4\times$ and saturating HBM2e bandwidth.
\item \textbf{Split over fused kernels} (fusion, \texttt{argmax},
  \texttt{sum}): two lean passes at full occupancy beat one
  register-hungry fused pass on HIP.
\item \textbf{Rust macro dispatch} (all kernels): one match-based macro
  routes each call to the right vendor backend, so HIP and CUDA diverge
  only where the hardware demands it.
\end{enumerate}
These are deliberately low-level and mechanical, which is the point:
they can be applied kernel-by-kernel, validated in isolation, and
carried across libraries.

\section{Benchmarking methodology}
\label{sec:bench}
Reliable microbenchmarks are notoriously easy to get wrong, so we drive
all measurements from a statistics-driven harness in Rust (the Criterion
framework~\cite{ref:criterion}), reporting the mean over $100$ samples per configuration
rather than a single timed run. Each kernel is exercised at three input
sizes chosen to stress the L1, L2 and L3/main-memory regimes
respectively, so that cache effects and the GPU launch floor are visible
rather than averaged away. The GPU measurements include the PCIe copy,
so the comparison is end-to-end rather than device-resident only.
Benchmark results and scripts are archived
publicly~\cite{ref:benchrepo}.

The primary test node, \texttt{della-milan}, is summarized in
Table~\ref{tab:node}. It pairs two AMD EPYC~7763 (Milan / Zen~3) sockets
with two AMD Instinct MI210 (CDNA2, \texttt{gfx90a}~\cite{ref:cdna2}) accelerators, so
that a fair CPU-versus-GPU comparison can be made on the same host, and
so that the GPU path exercises the discrete, non-unified-memory
configuration typical of current deployments. A subset of the kernels
is additionally cross-checked on Apple~M4 Metal hardware, confirming
that the engine's portability is not specific to a single non-NVIDIA
target.

\begin{table}[t]
\centering
\caption{The \texttt{della-milan} benchmark node. CPU and GPU share the
  same host, enabling an end-to-end comparison that includes the PCIe
  transfer.}
\label{tab:node}
\small
\begin{tabular}{@{}p{0.16\linewidth}p{0.38\linewidth}p{0.34\linewidth}@{}}
\toprule
 & \textbf{CPU} & \textbf{GPU} \\
\midrule
Device &
  $2\times$ AMD EPYC~7763 (Milan / Zen~3) &
  $2\times$ AMD Instinct MI210 (CDNA2, \texttt{gfx90a}) \\[2pt]
Parallelism &
  $128$ cores total ($64$/socket, no HT), $3.53$~GHz boost, AVX2 &
  $104$ CUs, $6\,656$ shaders \\[2pt]
Memory &
  $\sim$1~TB DDR4, L2 $64$~MiB, L3 $512$~MiB, $8$ NUMA nodes &
  $64$~GB HBM2e, $\sim$1.6~TB/s \\[2pt]
Notes &
  glibc, ROCm host &
  PCIe (discrete, no unified memory) \\
\bottomrule
\end{tabular}
\end{table}

\section{Results}
\label{sec:results}
\subsection{GPU versus CPU on irregular kernels}
\begin{figure}[t]
\centering
\includegraphics[width=0.64\linewidth]{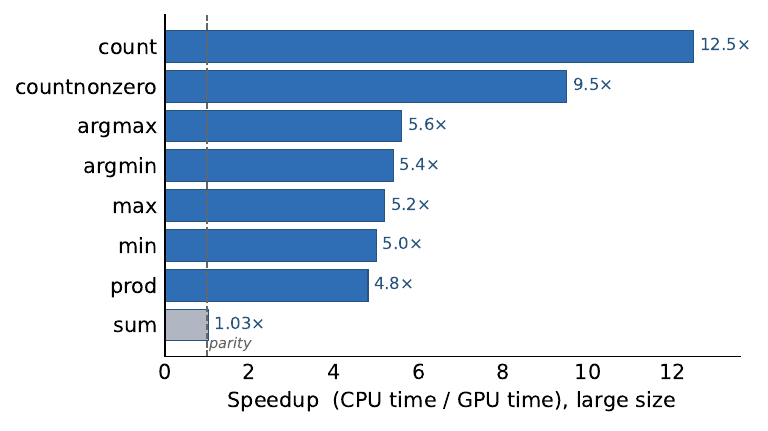}
\caption{Per-kernel speedup of the HIP backend on $2\times$ MI210
  against $128$ EPYC~7763 cores at the large input size (higher is
  better; the dashed line marks parity). Write-dominated kernels win
  most; the bandwidth-bound \texttt{sum} reaches only parity because
  AVX2 already saturates CPU memory bandwidth.}
\label{fig:speedup}
\end{figure}

Figure~\ref{fig:speedup} shows per-kernel GPU speedups over the full
$128$-core CPU at the large input size. The kernels fall into clear
tiers. \texttt{count} achieves the best result, $12.5\times$: it is a
write-only pattern with no arithmetic, so the GPU issues a single read
per segment and is limited only by memory throughput.
\texttt{countnonzero}, a predicated write, follows at $9.5\times$. The
argument- and value-reductions---\texttt{argmax}, \texttt{argmin},
\texttt{max}, \texttt{min} and \texttt{prod}---are compare-reduce,
memory-bound kernels and land in the $4.8$--$5.6\times$ band. At the
other extreme, \texttt{sum} ties the CPU at $1.03\times$: AVX2
auto-vectorization on the EPYC reaches full memory bandwidth at the
$1$M-element size, erasing the GPU's edge for this bandwidth-bound
reduction.

The behaviour at small sizes is governed by a fixed GPU launch floor of
roughly $35$--$36\,\mu$s. Below input sizes of a few thousand elements
this overhead dominates and the CPU wins; the crossover point sits
between $1$K and $64$K elements, above which the GPU wins cleanly. This
is not a defect but a design boundary: it tells an analysis framework
where it is worth dispatching to the accelerator and where it is not. It
is also why the \texttt{argsort} kernel is a striking case---the GPU
beats all $128$ CPU cores even on relatively small lists, and even with
the PCIe copy included, because the sorting work per element is high
enough to amortize both the launch floor and the transfer.

\subsection{Rust versus C++ CPU kernels}
\begin{figure}[t]
\centering
\includegraphics[width=0.64\linewidth]{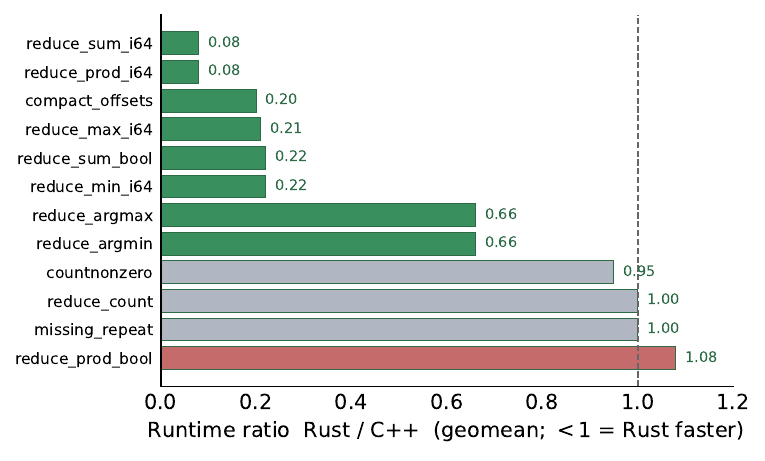}
\caption{Rust-versus-C++ CPU kernel runtime ratio (geometric mean over
  three input sizes; lower is better, dashed line marks parity). Green
  bars are Rust wins, grey is parity, red is a regression. Overall
  geometric-mean ratio $0.37\times$ across twelve kernels.}
\label{fig:rustcpp}
\end{figure}

The foundry model requires that the Rust kernels be at least as good as
the C++ kernels they are meant to replace, on the CPU baseline, before
any GPU discussion is meaningful. Figure~\ref{fig:rustcpp} reports the
per-kernel runtime ratio of the Rust implementation to the incumbent
C++ implementation (Criterion mean over $100$ samples; values below
$1.0$ mean Rust is faster). The Rust kernels are dramatically faster on
the reductions that vectorize well (\texttt{reduce\_sum\_i64} at
$82.9\,\mu$s versus $1.64$~ms, and \texttt{reduce\_prod\_i64}, both near
$0.08\times$---an order of magnitude), comfortably faster on offset
compaction and min/max ($\sim$0.20$\times$), and at parity on the
remaining kernels, with a single kernel (\texttt{prod\_bool}) marginally
slower at the smallest size but recovering to parity at scale. The
overall geometric-mean runtime ratio is $0.37\times$ across twelve
kernels: the Rust core is, on aggregate, comfortably faster than the
C++ baseline while adding compile-time memory-safety guarantees.

\subsection{Profiling summary}
Taken together, the measurements draw a consistent picture: a hard GPU
launch floor near $35\,\mu$s sets the small-size crossover; write-domin\-ated
kernels (\texttt{count} at $12.5\times$, \texttt{countnonzero} at
$9.5\times$) extract the largest speedups because they are limited only
by memory throughput; compare-reduce kernels
(\texttt{argmax}/\texttt{argmin}/\texttt{max}/\texttt{min}/\texttt{prod})
occupy a $4.8$--$5.6\times$ middle tier; and bandwidth-bound reductions
such as \texttt{sum} reach only parity once AVX2 saturates CPU
bandwidth. The lesson for a portable engine is that the \emph{kernel's
memory-access character}, not the vendor label, predicts where
acceleration pays off---once the HIP-specific patterns of
Sect.~\ref{sec:patterns} have removed the artificial penalty of a
na\"ive port.

\section{Roadmap and future work}
\label{sec:roadmap}
Four directions structure the work ahead. \textbf{Kernel completeness}:
extend beyond reductions so that carry, flatten, mask and sort join
reduce across every numeric type on CPU, Metal and HIP.
\textbf{Adaptive performance}: replace hand-tuned launch constants with
profile-guided block- and tile-selection chosen at first run per kernel
and per GPU architecture and then cached, with HIP auto-tuning validated
on MI300A. \textbf{Unified GPU backend}: collapse the current per-target
kernel code behind a single \texttt{GpuBackend} trait spanning Metal,
HIP and CUDA, so there is one dispatch path and zero per-target
boilerplate in kernel code---the macro dispatch of
Sect.~\ref{sec:rust} is the first step toward this. \textbf{Ecosystem
integration}: provide native columnar interoperability with RDataFrame
and coffea so that \emph{rawkward} arrays are usable in HEP pipelines
without copy or conversion, and validate the whole stack on MI300A-class
clusters. Looking further ahead, the CDNA4 MI350 generation and its
$288$~GB of HBM3E memory will relax the memory-capacity constraints that
currently shape kernel tiling.

\section{Conclusion}
\label{sec:conclusion}
HIP kernels for irregular data are not trivial: a direct port of an
optimized CUDA kernel can run $5$--$10\times$ slower, and the reason is
architectural rather than a matter of toolchain maturity. But the gap is
closable. With a small, reusable set of patterns---loop flattening,
$128$-bit vectorized loads, splitting kernels rather than fusing them,
and a Rust macro that dispatches each call to the right vendor
backend---the HIP kernels match their CUDA counterparts and unlock real
performance portability. On a two-socket MI210 node the resulting
backend delivers up to $12.5\times$ over $128$ CPU cores where the
memory pattern favours the GPU, and reaches parity precisely where the
CPU already saturates its bandwidth, while the underlying Rust kernels
beat the incumbent C++ kernels by a geometric-mean factor of
$0.37\times$ on the CPU baseline. The same Rust API drives both vendors,
with the hardware-specific choices hidden behind the dispatch layer. For
the HL-LHC era, in which AMD accelerators are no longer the exception,
that combination---one API, parity performance, memory safety by
construction---is a practical route toward a truly hardware-agnostic HEP
analysis stack.

\section*{Acknowledgements}
This work was supported by the U.S.\ National Science Foundation (NSF)
under Cooperative Agreements OAC-1450377, OAC-1836650, OAC-2103945,
PHY-2121686 and PHY-2323298, and carried out within the Institute for
Research and Innovation in Software for High Energy Physics (IRIS-HEP).
The authors thank the IRIS-HEP and Awkward Array communities, and
acknowledge the Princeton Research Computing \texttt{della-milan}
resources used for the benchmarks.

\footnotesize
\setlength{\itemsep}{-1.5pt}


\begin{thebibliography}{10}
\setlength{\itemsep}{-1.5pt}
\bibitem{ref:letts}
  J.~Letts, \textit{WLCG Technical Evolution: Preparing for HL-LHC},
  these proceedings

\bibitem{ref:rembser}
  J.~Rembser, \textit{Particle physics data analysis on the GPU},
  these proceedings

\bibitem{ref:mohrman}
  K.A.~Mohrman, \textit{GPU acceleration of end-user analyses at the
  LHC}, these proceedings

\bibitem{ref:awkward}
  J.~Pivarski et al., \textit{Awkward Array}, Zenodo (2018),
  \url{https://github.com/scikit-hep/awkward}

\bibitem{ref:naumchyk-cuda}
  M.~Naumchyk et al., \textit{CUDA Acceleration of Awkward Array Using
  Python CCCL}, these proceedings

\bibitem{ref:naumchyk-coffea}
  M.~Naumchyk et al., \textit{Coffea schema modifications for GPU
  backends}, these proceedings

\bibitem{ref:benchrepo}
  I.~Osborne, \textit{Bridging the Vendor Gap: benchmark data and
  scripts}, GitHub (2026),\\
  \url{https://github.com/ianna/CHEP2026-Bridging-the-Vendor-Gap}

\bibitem{ref:rocm}
  Advanced Micro Devices, \textit{ROCm/HIP Programming Guide} (2025),
  \url{https://rocm.docs.amd.com}

\bibitem{ref:cdna2}
  Advanced Micro Devices, \textit{AMD CDNA2 Architecture White Paper}
  (2022)

\bibitem{ref:criterion}
  B.~Heisler et al., \textit{Criterion.rs: Statistics-driven
  benchmarking for Rust} (2024),
  \url{https://github.com/bheisler/criterion.rs}
\end{thebibliography}
\end{document}